\documentclass[aps,prl,onecolumn,superscriptaddress,showpacs]{revtex4}
\usepackage{graphicx,amsmath}
\usepackage{epsfig}
\usepackage{gensymb}
\begin{document}

% Use the \preprint command to place your local institutional report
% number in the upper righthand corner of the title page in preprint mode.
% Multiple \preprint commands are allowed.
% Use the 'preprintnumbers' class option to override journal defaults
% to display numbers if necessary
%\preprint{}

\title{Characteristics of Cherenkov Radiation in Naturally Occuring Ice}
\author{R.E. Mikkelsen}
\affiliation{Department of Physics and Astronomy, Aarhus University, Denmark}
\author{T. Poulsen}
\affiliation{Department of Physics and Astronomy, Aarhus University, Denmark}
\author{U.I. Uggerh{\o}j}
\affiliation{Department of Physics and Astronomy, Aarhus University, Denmark}
\author{S.R. Klein}
\affiliation{Lawrence Berkeley National Laboratory, Berkeley CA 94720, USA}
\affiliation{Physics Dept. University of California, Berkeley, CA 94720, USA}

\date{\today}
\begin{abstract}
We revisit the theory of Cherenkov radiation in uniaxial crystals. 
Historically, a number of flawed attempts have been made at explaining this radiation phenomenon and a consistent error-free description is nowhere available. 
We apply our calculation to a large modern day telescope - IceCube. 
Being located at the Antarctica, this detector makes use of the naturally occuring ice as a medium to generate Cherenkov radiation.
However, due to the high pressure at the depth of the detector site, large volumes of hexagonal ice crystals are formed.
We calculate how this affects the Cherenkov radiation yield and angular dependence. 
We conclude that the effect is small, at most about a percent, and would only be relevant in future high precision instruments like e.g. Precision IceCube Next Generation Upgrade (PINGU).
For radio-Cherenkov experiments which use the presence of a clear Cherenkov cone to determine the arrival direction, any variation in emission angle will directly and linearly translate into a change in apparent neutrino direction.
In closing, we also describe a simple experiment to test this formalism, and calculate the impact of anisotropy on light-yields from lead tungstate crystals as used, for example, in the CMS calorimeter at the CERN LHC. 

\end{abstract}

%\maketitle must follow title, authors, abstract, \pacs, and \keywords
\maketitle

\section{Introduction}
Cherenkov radiation is a well-understood phenomenon related to the passage of charged, ultra relativistic particles through a dielectric medium. Since its discovery by P. A. Cherenkov in 1934~\cite{Cherenkov}, it has been studied both experimentally and theoretically with such success that Cherenkov radiation is now routinely used in applications such as Cherenkov particle counters; study of biomolecules and in astronomical observatories.  One particularly interesting application is to search for energetic neutrinos from cosmic sources.  Because of their low flux and small interaction cross-sections, this requires very large detectors, with volume of order 1 km$^3$.
%Experiments like the Pierre Auger observatory, Super-Kamiokande, Sudbury Neutrino Observatory and IceCube~\cite{IceCube} detect Cherenkov radiation to study particles of extraterrestrial origin interacting in or near the detector. [n.b. Super-K and SNO don't primarily search for neutrinos from cosmic sources; and Pierre Auger observe fluorescence, not Cherenkov radiation. 
%Because of the rarity of the most energetic cosmic particles, there is a persistent demand for larger ground based observatories; in the case of neutrinos, the low cross sections only adds to this trend.
To build a detector with volume of order 1 km$^3$ requires naturally occuring water or ice as the Cherenkov medium.   Currently, the largest detector is IceCube \cite{IceCube}, located at the South Pole.  It observes the Cherenkov radiation from the charged particles produced in neutrino-induced showers.  To understand this radiation, it is necessary to understand how charged particles radiate photons in the Antarctic ice.   This ice consists of  hexagonal ice crystals that are oriented in the same direction\cite{ice,ice2}. This orientation leads to an anisotropy, and the Cherenkov radiation may depend on the direction of the ice orientation.  This is of particular interest because IceCube has already observed an anistropy in the ice, believed to be due to the scattering depending on the azimuthal direction the photon follows through the medium \cite{ICice}.  Any Cherenkov production anisotropies may be confused with this scattering anisotropy and the presence of a directional Cherenkov anisotropy can also affect neutrino directional and energy reconstructions.  This is particularly important for the next-generation PINGU detector, which will need to reach very low levels of systematic error to be able to determine the neutrino mass hierarchy \cite{Aartsen:2014oha}.

The first theory for Cherenkov radiation in an isotropic medium was produced by I. Y Tamm and I. M Frank in 1937~\cite{FrankTheTank}. 
Since then a number of flawed attempts have been made to describe the same process in an anisotropic medium where the particle propagation direction becomes important for the Cherenkov emission.
In 1956, V. E. Pafomov did a calculation~\cite{Pafomov} that reproduced the correct emission angles and described the intensity distribution, but failed to give the correct result when the calculation was applied to an isotropic material.
Then in 1960, C. Muzicar~\cite{Muzicar} obtained a similar result for the emission angles but a different dependence of the number of photons emitted on the propagation direction of the emitting particle with respect to the symmetry axis of the medium. 
An experiment carried out by D. Gf\"oller \cite{Gfoller} in the following year was in support of Pafomovs result.
The first fully convincing calculation for the Cherenkov photon emission from an anisotropic material was given in a 1997 paper~\cite{Delbart} by A. Delbart, J. Derr\'e and R. Chipaux; despite the succesful calculation, however, the paper includes a number of typing errors in the central equations for the Cherenkov photon yield. 

In this paper, we provide a corrected version of the description of the Cherenkov radiation in a uniaxial medium as published by Delbart et. \textit{al}. We then proceed to calculate some energy spectra and discuss the possible implications for IceCube and whether this should be implemented in the in-ice propagation codes. 
Lastly, we prescribe a procedure to test the present calculations in a small-scale experiment, and results relevant for high-energy calorimetry.
\section{Theory of Cherenkov radiation in uniaxial optical materials}
{\label{sec:theory}}
Consider a relativistic charged particle with velocity $\beta c$ moving in an isotropic medium of refractive index, $n$.
The particle emits Cherenkov radiation if the condition $\beta n > 1$ is fulfilled; that is, only for relativistic particle speeds; throughout this paper, we take as input a relativistic charged particle with $\beta=1$.
The radiation is emitted at a characteristic angle, $\cos\theta_{\textrm{C}} = (\beta n)^{-1}$, with respect to the propagation direction of the emitting particle. 
Contrary to other relativistic radiation phenomena like synchrotron radiation and bremsstrahlung, this angle 
can be quite large:  $40^{\degree}$ for ice, and $70^{\degree}$ for the birefringent material rutile; both of which will be discussed later. The emission angle is azimuthally symmetric in isotropic materials. 
If the material is anisotropic this symmetry may be broken, and  the refractive index may depend on the angle of the incoming particle, complicating the radiation pattern.

Here, we study uniaxial materials where the optical axis defines a direction around which there is azimuthal symmetry. These materials have two different refractive indices. Light polarized along the optical axis experiences a refractive index, $n_e$ with $e$ for extraordinary; and light with a polarization perpendicular to the optical axis experiences a refractive index $n_o$, with $o$ for ordinary. These terms originate from optics theory and are applied to Cherenkov radiation only for anisotropic media. 
The ordinary wave denotes photons travelling along the optical axis. Their propagation does not depend on the photon polarization. In contrast,  the propagation properties of the extraordinary wave depend on their polarization.
Photons travelling at other angles experience a refractive index in between $n_o$ and $n_e$.
The relationship between $\mathbf{\hat{d}}$ which is a unit-vector in the direction of the displacement and $\mathbf{\hat{e}}$ which points in the direction of the electric field is $d_i = \Sigma \epsilon_{ik} e_k$.
This will be used along with the expression for the phase velocity, $v_p = c/ \sqrt{\mu \epsilon}$.
Combining these equations and using that $\mathbf{\hat{d}}$ is unitary gives $v_p^2= \mathbf{e}\cdot\mathbf{\hat{d}}\;c^2/ \mu$.

\begin{figure}[hbt]
\begin{center}
\includegraphics[width=.5\textwidth]{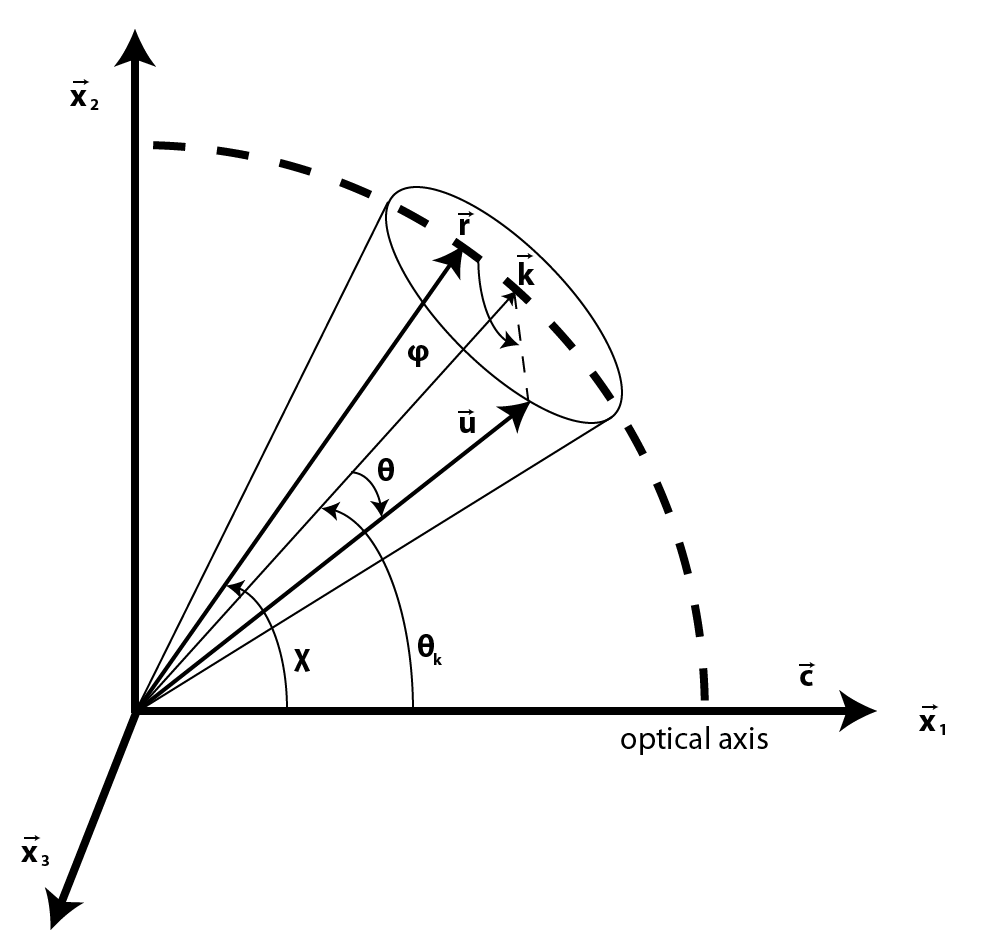} 
\caption{The coordinate system and the angles used in the calculation. The $\mathbf{x}_1$ axis is chosen to point in the direction of the optical axis. The angle $\chi$ between the incoming particle and the optical axis is in the plane spanned by $\mathbf{x}_1$ and $\mathbf{x}_2$. }
\label{fig:Geometry}
\end{center}
\end{figure}

We define the geometry as shown in Figure \ref{fig:Geometry}.   
Because of the symmetry around the optical axis, i.e. $\mathbf{x_1}$, the propagation direction of the emitting particle is defined by the single angle $\chi$, and written
\begin{align}
\mathbf{\hat{r}} = \begin{pmatrix} r_1 \\ r_2	\\ r_3	\end{pmatrix} = \begin{pmatrix} \cos\chi \\ \sin\chi \\ 0	\end{pmatrix},
\end{align}
such that the particle propagates in the $(\mathbf{x_1},\mathbf{x_2})$ plane.
When $\chi=0$, the optical axis is aligned along the direction of the radiating particle.
In this geometry the dielectric tensor is diagonal with $\epsilon_{11} = n_e^2$ and $\epsilon_{22} =\epsilon_{33} = n_o^2$.
The unit vector $\mathbf{\hat{k}}$ seen on Figure \ref{fig:Geometry} lies in the plane spanned by $\mathbf{\hat{r}}$ and the optical axis.
Rather than expressing a physical direction, $k_1$ and $k_2$ are free parameters that can be chosen such as to simplify the calculation of integrals in the following. 
The unit vector $\mathbf{\hat{u}}$ points in the direction of the Cherenkov wave phase propagation.
It is expressed in polar coordinates around $\mathbf{\hat{k}}$, as illustrated on Fig. \ref{fig:Geometry}:
\begin{align}
\mathbf{\hat{u}} =  \begin{pmatrix} u_1 \\ u_2 \\ u_3	\end{pmatrix} = \begin{pmatrix} k_1\cos(\theta) - k_2\sin(\theta)\cos(\phi) \\ k_2\cos(\theta) + k_1\sin(\theta)\cos(\phi) \\ \sin(\theta)\sin(\phi) \end{pmatrix}.
\end{align}
The vector $\mathbf{\hat{u}}$ points in the direction of photon propagation when there is no dispersion in the medium.
With these definitions, the expression for the differential number of photons emitted in a length interval $dl$ within an energy interval, $dE$, is~\cite{Ginzburg}
\begin{align}
\frac{d^2N}{dldE}=\frac{\alpha c^3}{2\pi \hbar \mu} \int_{0}^{4\pi}
\frac{(\mathbf{e}\cdot\mathbf{\hat{r}})^2}{v_p^4}\delta \left(\mathbf{\hat{r}}\cdot \mathbf{\hat{u}}-\frac{v_p}{c\beta}\right) d\Omega,
\label{eq:Ginzburg}
\end{align}
where $\alpha$ is the fine structure constant, $v_p$ is the phase velocity of the emitted wave, $\mathbf{e}$ is a vector in the direction of the electric field, and $\mu$ is the scalar magnetic permittivity of the medium - set to $1$ throughout here.
With the dielectric tensor as defined above, Eq. \ref{eq:Ginzburg} can be decomposed in two contributions representing the number of ordinary, $N^{(o)}$ and extraordinary $N^{(e)}$ photons respectively.

\subsection{The ordinary waves}

The following approach follows exactly \cite{Muzicar} and \cite{Delbart} but with a few corrections along the way. We use the notation from the latter to calculate the number of Cherenkov photons emitted.
For the ordinary wave, $\mathbf{\hat{r}}$ points along $\mathbf{\hat{k}}$ and the integration of Eq. \ref{eq:Ginzburg} over $\theta$ can be done by putting in the values of $\mathbf{e}$, $\mathbf{\hat{r}}$ and $\mathbf{\hat{u}}$.
We differentiate the result by $d\phi$ and obtain the triply differential  number of ordinary photons emitted~\footnote{In Ref.~\cite{Delbart}, $\cos(\chi)$ was written in the denominator rather than $\cos(\phi)$.}
\begin{align} 
\label{eq:N_o}
\frac{d^3N^{(o)}}{d l d E d\phi} = \frac{\alpha}{2\pi\hbar c} \times  \frac{\sin^2(\theta_{o})\sin^2(\chi)\sin^2(\phi)}{1-\left[\cos(\theta_{o})\cos(\chi)-\sin(\chi)\cos(\phi)\sin(\theta_{o})\right]^2}.
\end{align}
The argument of the Dirac delta function in (\ref{eq:Ginzburg}) determines the geometrical properties of the emitted photons. 
For the ordinary photons the phase velocity is $v_p^{(o)} = c/n_o$ and therefore $\mathbf{\hat{r}}\cdot\mathbf{\hat{u}}=1/n_o\beta=\cos (\theta _{(o)})$ which is the Cherenkov emission angle in an isotropic medium.
It is a general result that if the particle propagates along the optical axis, $\chi = 0$, no ordinary photons are emitted. 

\subsection{The extraordinary waves}
The calculation of the triply differential cross section of the number of emitted extraordinary photons follows a similar route as that for the ordinary photons. 
It is somewhat more complicated because now $\mathbf{\hat{r}} \neq \mathbf{\hat{k}}$.
However, $\mathbf{\hat{k}}$ is still a symmetry axis for the emission along $\mathbf{\hat{u}}$ although it turns out that the extraordinary cone has an elliptical shape.
Again, the first step is to evaluate the argument of the Dirac delta function which is $\mathbf{\hat{r}}\cdot\mathbf{\hat{u}} = v_p^{(e)}/c\beta$ for the extraordinary wave.
From the definition of the phase velocity, we get for the extraordinary wave
\begin{align}
v_p^{(e)} = c\sqrt{\cfrac{1}{n_e^2} + \left( \cfrac{1}{n_o^2} - \cfrac{1}{n_e^2} \right) u_1^2}.
\end{align}
Using this to solve the Dirac delta function to obtain the geometrical properties of the wave one finds that the choice 
\begin{align} 
\label{eq:k1k2}
k_1 = \frac{1}{\sqrt{2}}\sqrt{1+\frac{A}{\sqrt{A^2+4r_1^2r_2^2}}} \\
k_2 = \frac{1}{\sqrt{2}}\sqrt{1-\frac{A}{\sqrt{A^2+4r_1^2r_2^2}}},  
\end{align}
with 
\begin{align}
\label{eq:A}
A = \left( r_1^2-\cfrac{1}{n_o^2\beta^2} \right) - \left( r_2^2-\cfrac{1}{n_e^2\beta^2} \right),
\end{align}
simplifies the expression (\ref{eq:Ginzburg}) the most. 
In Appendix A, we show the two final steps needed to evaluate this expression for the extraordinary waves and discuss some errors presented in the results given in \cite{Muzicar,Delbart}.
The number of extraordinary photons emitted is 
\begin{align} 
\label{eq:N_e}
\frac{d^3N^{(e)}}{d l d E d\phi} = \frac{\alpha}{2\pi\hbar c} \frac{1}{n_o^2 \beta^2}\times 
\cfrac{-\cfrac{1}{\mathbf{\hat{u} \cdot \hat{r}} n_o^2\beta^2}+\cfrac{1}{1-u_1^2}\left(\mathbf{\hat{u} \cdot \hat{r}}\left(r_1^2\beta^2n_o^2-2\right)+\cfrac{1}{\beta^2n_o^2\mathbf{\hat{u} \cdot \hat{r}}} +2u_2r_2\right)}
{\sqrt{\left(R-Q\cos^2(\phi)\right)\left(P-Q\cos^2(\phi)\right)}},
\end{align}
where $R = \cfrac{1}{n_e^2\beta^2}$, $Q=\left(k_2r_1-k_1r_2 \right)^2 - \left(\cfrac{1}{n_o^2} - \cfrac{1}{n_e^2}\right)\cfrac{k_2^2}{\beta ^2}$ and $P=\left(k_1r_1+k_2r_2 \right)^2 - \left(\cfrac{1}{n_o^2} - \cfrac{1}{n_e^2}\right)\cfrac{k_1^2}{\beta ^2}$.

\subsection{Theory summary - observables}
In the previous sections we listed the corrected results for the number of ordinary and extraordinary photons differential in emission angle, energy and target thickness. 
When these contributions are integrated over angle and energy and subsequently added, we obtain the total number of photons emitted per pathlength $dl$. 
For the case when $\chi = 0^{\degree}$ and $\chi= 90^{\degree}$ we get respectively:

\begin{align}
N_\parallel^t &= \frac{\alpha}{\hbar c} \left(1- \frac{1}{n_0^2\beta^2}\right), \textrm{and}\\
N_\perp^t &= \frac{\alpha}{\hbar c} \left(1-  \frac{1}{n_0n_e\beta^2}\right).
\end{align}
When calculated in units of eV$^{-1}$ and mm$^{-1}$, $\frac{\alpha}{\hbar c} \approx 37$.
Later we will calculate the relative number $R_t = N_\perp^t /N_\parallel^t$, to compare the two most extreme scenarios. 
We will also assume that the refractive indices $n_0$ and $n_e$ are constant.
The overall change of the refractive index of ice from the top to the bottom of IceCube due to pressure and temperature variations was estimated~\cite{Price} to be $\Delta n = 0.002$.
Ref.~\cite{iceReview} gives a review of the dependence of the refractive index on the photon wavelength. 

\section{The Antarctic ice}
\begin{figure}[hbt]
\begin{center}
\includegraphics[width=1\textwidth]{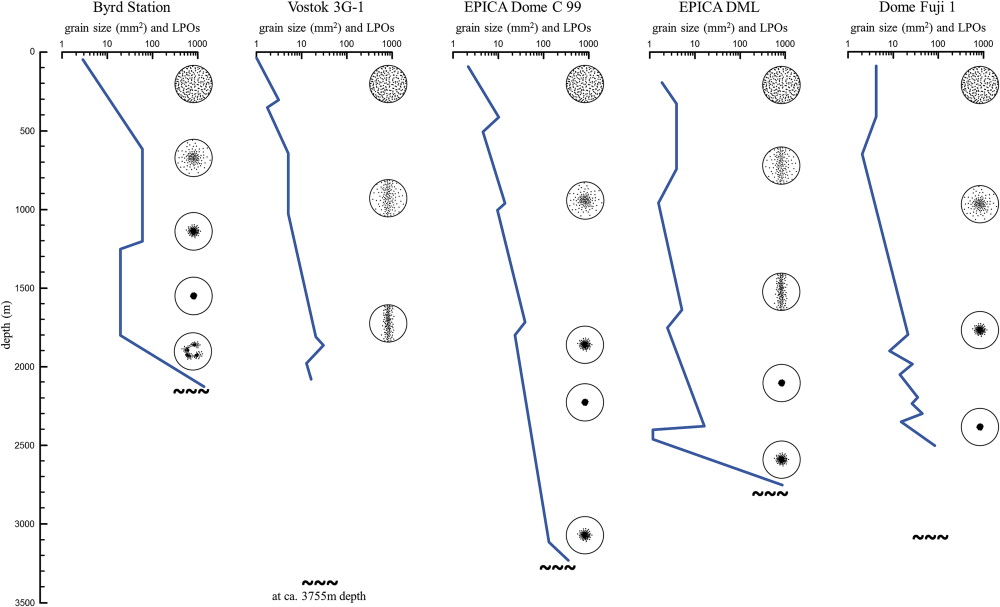} 
\caption{The blue line shows the grain size as a function of depth for Antarctic ice measured at five different locations. 
The circular plots called LPOs (see text for a definition) show the orientation of the ice crystals at various depths.
The dark tildes symbols represent the bedrock. 
The distance from the South Pole to Byrd station is 1100 km; to Vostok is 1250 km; to Dome C is 1765 km; to Kohnen station where EPICA DML was drilled is 1670 km; and to the Dome Fuji station is 1250 km. From \cite{ice}. }
\label{fig:AntarcticIce}
\end{center}
\end{figure}

We have seen in the theory section that the birefringence of crystals affects the emission of Cherenkov radiation.  This will be relevant for IceCube if all or most of the crystals in which radiation is emitted point in the same direction. 
Figure \ref{fig:AntarcticIce} summarizes from \cite{ice} the size and orientation of the Antarctic ice crystallites measured at five different locations. 
The circular plots are called LPO's for lattice-preferred orientation which shows the crystallographic orientations of the crystallites composing the sample, in stereographic projection; see \cite{LPO} for a textbook definition. 
They show the degree to which the individual crystals in the ice point in the same direction. 
At all five drilling locations, the ice is randomly oriented near the surface and more and more aligned deeper down, except when just next to the bedrock.  In all five samples, at the depths of the IceCube optical sensors  {\it i.e.} $1450-2450$ m, the ice crystals are not randomly oriented but rather have a high degree of common orientation.  Generally, the optical axis tends to rotate towards an axis of compression.  At large depths the ice crystals are no longer randomly oriented as at the surface, but have a preferred direction which depends on the flow history \cite{Dorte}.

No such measurement has been performed at the Geographic South Pole where IceCube is located, and the IceCube dust logger \cite{dustlogger} is insensitive to any azimuthal dependence of the ice properties.    However, the IceCube optical modules are equipped with LED flashers which can send signals to other IceCube optical modules. 
Using this system, the IceCube collaboration has observed a significant anisotropy in the scattering of light at the detector site~\cite{IceCubeIce}. 
The anisotropy is oriented along the direction of ice flow. % - which is about $10$ m per year.
It is typically attributed to dust grains in the ice which should be aligned by the pressure gradient in the ice \cite{Bay}, but may have other origins such as the effect discussed here.

\section{Results}

\subsection{Relevance for IceCube}
\begin{figure}[hbt]
\begin{center}
\includegraphics[width=.5\textwidth]{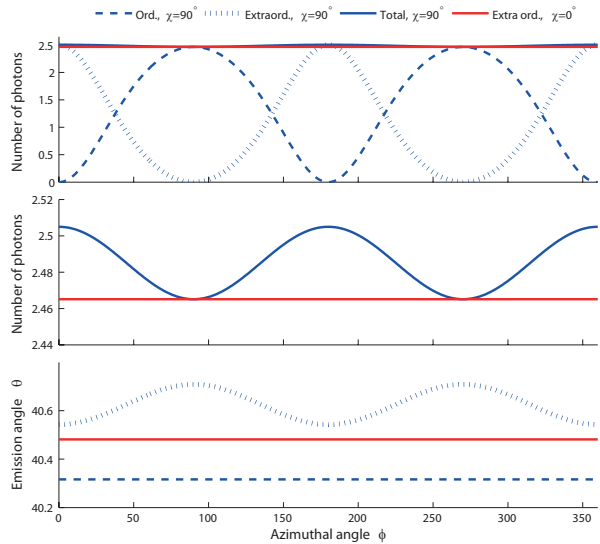} 
\caption{Properties of Cherenkov radiation emitted in a hexagonal ice crystal. Top: The number of photons, $\frac{d^3N}{dldEd\phi}$, as a function of azimuthal angle $\phi$. The blue lines are for $\chi = 90^{\degree}$ and the red curve is for $\chi = 0^{\degree}$ - see legend for details. Middle: A zoom of the above figure so that the variation in total emitted photons emitted when $\chi = 90^{\degree}$ can be seen. Bottom: Cherenkov radiation emission angle as a function of $\phi$.}
\label{fig:HexagonalIce}
\end{center}
\end{figure}

\begin{figure}[hbt]
\begin{center}
\includegraphics[width=.5\textwidth]{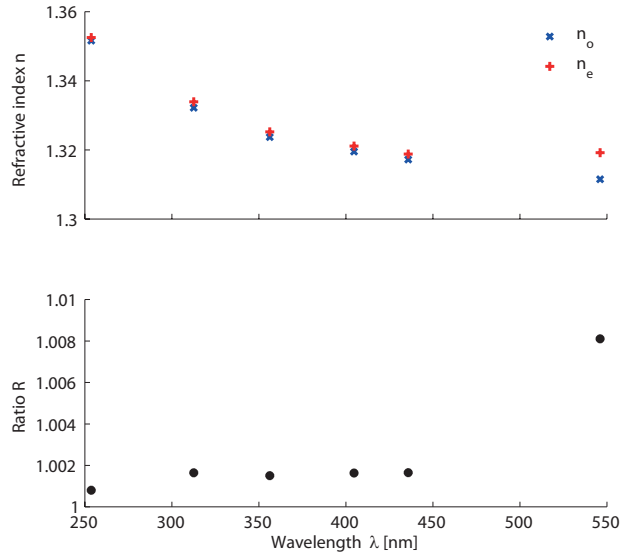} 
\caption{Top: The ordinary and extraordinary refractive indices of ice for different photon wavelengths~\cite{Japan}. 
The crosses represent data measured at an ice temperature of $-27.5^{\degree}$C; presented without error bars as in the original paper. 
The extraordinary refractive index at the largest wavelength of $546$ nm differs somewhat from the trend. 
Bottom: The ratio $R_t = N_\perp^t /N_\parallel^t$ calculated using the above values. }
\label{fig:wavelength}
\end{center}
\end{figure}

We now apply the results presented in the theory section to Antarctic ice, with a focus on the extreme cases with $\chi = 0^{\degree}$ and $\chi = 90^{\degree}$.
Figure \ref{fig:HexagonalIce} shows the number of emitted photons per azimuthal angle, $d\phi$, energy, $dE$, and path length $dl$, as a function of azimuthal angle, $\phi$.
We use values from Figure \ref{fig:wavelength} for $n_o= 1.3115$ and $n_e=1.3192$ given by \cite{Japan} for an ice temperature of $-27.5^{\degree}$ C in agreement with the standard reference for the optical properties of ice \cite{IceIndices}. The temperature of the ice in IceCube varies between $-32$ at the top to $-9^{\degree}$ C at the lowest elevation optical detector modules~\cite{Lutz,IceCubeData}. 
The refractive indices also depend on photon wavelength.
To study the extreme case, we use a value of $546$ nm where $n_o$ and $n_e$ differ the most.
The detector optical modules used in IceCube are most sensitive at $390$ nm \cite{MaxEfficiency}, but at short wavelengths, light scatters much more than at longer wavelengths, so travels a shorter distance from its source.
The blue curves in Figure \ref{fig:HexagonalIce} pertain to the case $\chi = 90^{\degree}$; the full drawn curve being the sum of the ordinary photons (dashed line) and the extraordinary (dotted line). 
For the case $\chi = 0^{\degree}$, no ordinary photons are emitted and the solid red line only stems from the contribution of extraordinary photons. 
The difference between the solid blue and red curves is the net difference in Cherenkov radiation yield depending on the particle propagation direction to the optical axis.
The lower plot on Figure \ref{fig:HexagonalIce} shows how the Cherenkov emission angle $\theta$ depends on the azimuthal angle $\phi$. 
When $\chi=90$, the ordinary photons are emitted at a constant angle $\theta_{o}=40.32^{\degree}$, while the emission angle of extraordinary photons varies by $0.17^{\degree}$ depending on the $\phi$ angle; with peaks located at $90^{\degree}$ and $270^{\degree}$.
The solid red curve shows the constant emission angle of $\theta = 40.48^{\degree}$ of the extraordinary photons which is the sole contribution when $\chi = 0^{\degree}$.
Lastly, Figure \ref{fig:HexagonalIce} shows that the hexagonal nature of ice affects the number of Cherenkov photons emitted by up to $0.43\%$ depending on the angle $\chi$ of propagation of the radiating particle.

The dependence on wavelength of the index of refraction influences the results.
Data for both $n_o$ and $n_e$ is available in a range of wavelengths starting from $250$ to $550$ nm~\cite{Japan}.
In both cases, the index of refraction is increasing with decreasing wavelength. 
For the available data points, we have recalculated the ratio $R$ as a function of wavelength, see Figure \ref{fig:wavelength}.
The top figure shows for increasing wavelengths both the ordinary and extraordinary refractive indices decrease. 
The bottom figure shows that the ratio $R$ is relatively stable between wavelengths from about $250$ up to about $450$ nm. 
The single datapoint at a wavelength of $550$ nm gives an $R$ value which is three to four times larger; this is also where the difference between $n_o$ and $n_e$ is largest according to the top figure. 

The effect of anisotropic Cherenkov emission on neutrino detection depends on the observation channel.  
For the most energetic event yet seen by IceCube, with an observed energy of 2.6 PeV, the collaboration reported a preliminary angular uncertainty of 0.26 degrees \cite{ATEL}, comparable to these variations.  However, this was a muon track; the excellent angular resolution came from the 1 km long lever arm, and it is unlikely that small variations in emission would affect this reconstruction.  This also holds for less energetic muons.

In IceCube, $\nu_e$ are observed through their electromagnetic showers, where the direction is determined by observing the effects of the Cherenkov cone.   However,  in IceCube most photons scatter before they are detected, so the Cherenkov cone is largely washed out, and the angular uncertainty for these events is at least 10 to 15 degrees, increasing at lower energies \cite{Aartsen:2014gkd}.  It seems unlikely that the variation in Cherenkov angle will have an effect on current resolution; especially considering that uncertainties due to light propagation in the ice are more significant with an effect of order 10 \% in IceCube.  
However, it could be important with smaller detectors with less scattering ({\it e. g.} PINGU), and improved reconstruction algorithms with better angular resolution. The amplitude variation of 0.3\% is smaller than the energy uncertainty of 15 \% for IceCube (above 10 TeV), so does not seem to be significant~\cite{ICenergyResolution}.

\subsection{Anisotropic radio Cherenkov emission}
A few other experiments use the Antarctic ice as a Cherenkov medium.  The Askaryan Radio Array (ARA) ~\cite{ARA}  and the Antarctic Ross Ice-Shelf ANtenna Neutrino Array (ARIANNA)~\cite{ARIANNA1} are proposed experiments to study extremely rare cosmic neutrinos with energy above $\sim 10^{17}$ eV.   Both designs take advantage of the Askaryan effect~\cite{Askaryan}: The interaction of a cosmic neutrino with a nucleus near the detector causes an extended particle shower with a net negative charge due to annihilation of shower positrons with electrons in the medium through which the shower propagates.  The propagation of this net charge leads to the emission of radiowaves - the Askaryan effect. 

ARA will consist of an array of multiple measuring stations distributed over roughly $200$ m$^2$; $16$ test stations are already in place, taking data.  ARIANNA is a proposed array of multiple measurement stations distributed over roughly $900$ km$^2$ on the Ross ice shelf in Antarctica; 7 test stations are already  taking data~\cite{ARIANNA}.
ARIANNA is located on a floating ice-shelf with a thickness of $\sim 580$ m and it aims at performing both direct and  indirect measurements of  high energy cosmic neutrinos; in indirect measurements, the radio waves reflect off the ice-water interface before reaching the detectors. 
The ice sheet at the ARIANNA site was largely formed on the Antarctic plateau and gradually pushed North over a time scale of 100,000 years, going through gaps in the trans-Antarctic mountains onto the Ross sea. Anisotropies may have formed before, during and/or after the course of this transport.

These experiments need to  understand how radio waves propagate through Antarctic ice.  Fortunately, partly because radar is used to survey Antarctic ice and the underlying rock, a fair amount is known about the ice.    The refractive indices depend on depth.  Below the firn $R_t = N_\perp^t /N_\parallel^t=0.9975$ at $9.7$ GHz \cite{RadioNeNo}, similar to the value obtained at $39$ GHz~\cite{RadioNeNo2}.
The results reported on the two refractive indices compare well with the standard value $n=3.18$~\cite{n318} at radio frequencies referred to in ARIANNA papers. 

Ice crystal orientations are believed to play a significant role, and birefringence due to oriented ice crystals is significant.  
Refs.~\cite{AnisotropyAndFlow,AnisotropyAndFlow2} show that the scattering of radiowaves in Antarctic ice depends on the angle between the radio polarization and the direction of ice flow.   So, possible anisotropy in Cherenkov radiation must be considered. 
 
Greenland also has glaciers which could host a high-energy neutrino detector; this would provide Northern 
hemisphere coverage. One group has measured radio propagation near the Summit Station in Greenland~\cite{Greenland}.
They find radio attenuation properties that are, after accounting for the warmer temperatures, similar to those measured at the South Pole.  Likewise, studies of ice crystal orientation seem to be similar.  A plot similar to the present Figure \ref{fig:AntarcticIce}, but for Greenland, is available in~\cite{ice}.

Radio waves do not scatter significantly in the ice, so radio-detection experiments directly observe the Cherenkov cone; in fact measurements of the radio spectrum is used for directional reconstruction, by determining how far off the Cherenkov cone the observer is. So, any alteration of the Cherenkov cone is more important for radio-detection than for optical experiments; the change in Cherenkov angle translates fairly directly into angular uncertainty and hence the leads to a change in apparent neutrino direction.  The fractional changes in index of refraction between the two extreme directions are somewhat larger than at optical frequencies, at least at atmospheric pressure; the pressure dependence of this difference has not been studied.  
In the top $150$ m, the refractive index of the Antarctic ice changes from $1.35$ to $1.78$ i.e. by $32\%$ for radio waves \cite{nOnDepth,nOnDepth2}.
The pressure dependence of the refractive indices has not been studied for hexagonal ice, so it is unknown if the difference between $n_o$ and $n_e$ is constant with increasing pressure.
\subsection{Calculations for extreme cases, and a possible experiment to test the theory}

Here, we calculate the angular emission spectra for Cherenkov photons  in a selection of anisotropic media. 
Table \ref{tbl:Crystals} shows three materials chosen for their large anisotropy: rutile (TiO$_2$), calcium carbonate (CaCO$_3$), sodium nitrate (NaNO$_3$), plus hexagonal ice (H$_2$O I$_h$) crystals.
% The three first materials are chosen because of their large difference between $n_e$ and $n_o$, see . 
\begin{table}[ht]
\centering
\begin{tabular}{l l l l c}
Crystal			& $n_o$		& $n_e$\ \ \ \ 		& $R$ \ \ \ 			& Max ang. diff. 	\\
\toprule
TiO$_2$			& 2.616		& 2.903		& 1.017 		& 2.32$^{\degree}$	\\
CaCO$_3$ & 1.6584	& 1.4864\  \ 	& 0.9339 			& 5.20$^{\degree}$	\\
NaNO$_3$	& 1.5854		& 1.3369\ \ 		& 0.8433 	& 9.40$^{\degree}$	\\
H$_2$O I$_h$	& 1.309		& 1.313		& 1.003 	& 0.39$^{\degree}$	\\
\end{tabular}
\caption{Properties of three uniaxial crystals %\citep{Hecht} 
plus H$_2$O I$_h$
 and a summary of the influence on the Cherenkov radiation originating from these materials. The rightmost column shows the maximum difference between the emission angle for ordinary and extraordinary photons when $\chi = 90^{\degree}$.}
\label{tbl:Crystals}
\end{table}

\begin{figure}[hbt]
\begin{center}
\includegraphics[width=1\textwidth]{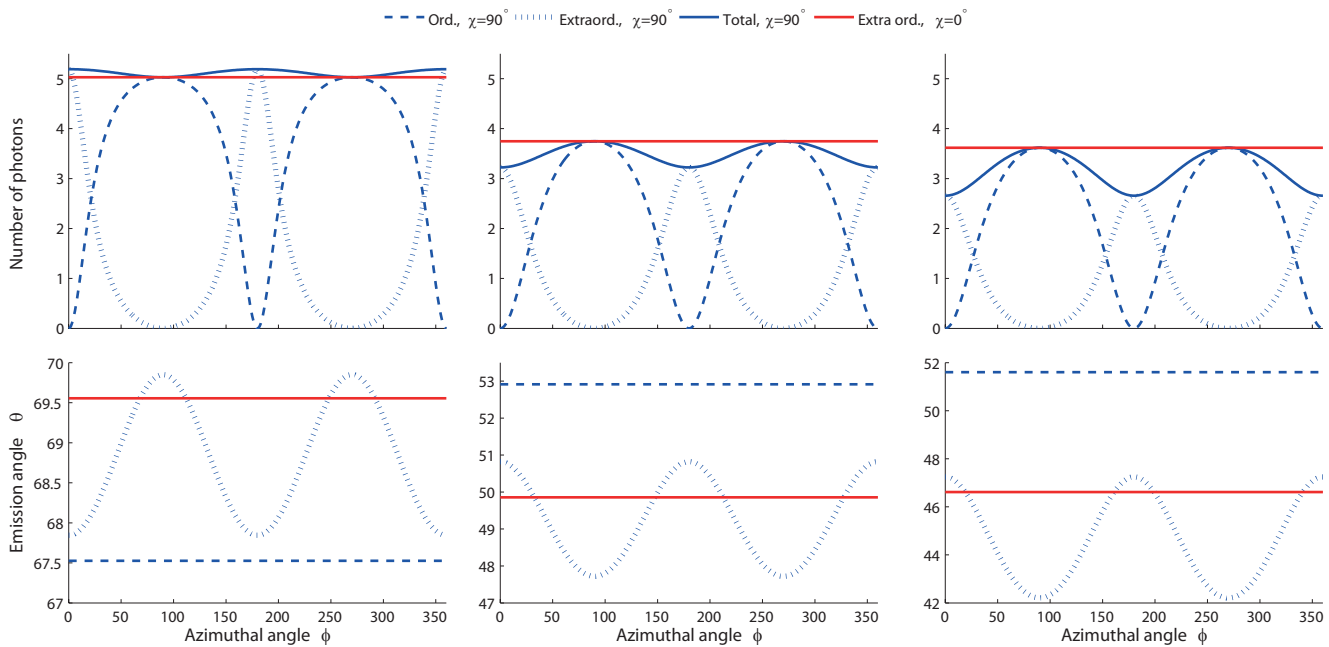} 
\caption{Properties of Cherenkov radiation emitted in three different crystals. For legend, see Figure \ref{fig:HexagonalIce} Top: Number of photons, $\cfrac{d^3N}{dldEd\phi}$, as a function of azimuthal angle $\phi$. Bottom: Emission angle as a function of $\phi$. Left: Rutile, mid: Calcium carbonate, right: sodium nitrate.}
\label{fig:Crystals}
\end{center}
\end{figure}

The results of the calculations are summarized for the three crystals on Figure \ref{fig:Crystals}.
We apply the same legends and distinctions between top and bottom plots as on Figure \ref{fig:HexagonalIce}.
As with hexagonal ice, no ordinary photons are emitted for $\chi =0$, so there is only one red curve on all the plots.
Oscillatory behavior for the ordinary, extraordinary and the sum of Cherenkov photons is visible for the three crystals. 
The maximum differences in emission angle, $\theta$, are summarized in table \ref{tbl:Crystals} along with the relative number of emitted photons, $R$.
For calcium carbonate and sodium nitrate, $n_e < n_o$ and hence the total emitted number of photons is larger when $\chi = 0$.

\begin{figure}[hbt]
\begin{center}
\includegraphics[width=.5\textwidth]{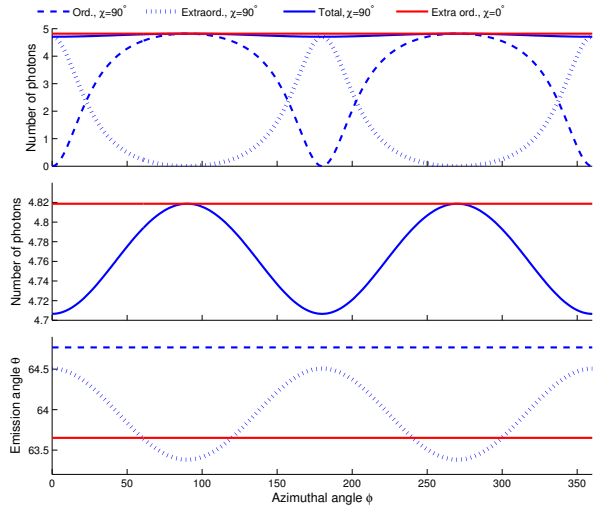} 
\caption{Properties of Cherenkov radiation emitted in a lead tungstate crystal. Top: The number of photons, $\frac{d^3N}{dldEd\phi}$, as a function of azimuthal angle $\phi$. The blue lines are for $\chi = 90^{\degree}$ and the red curve is for $\chi = 0^{\degree}$ - see legend for details. Middle: A zoom of the above figure so that the variation in total emitted photons emitted when $\chi = 90^{\degree}$ can be seen. Bottom: Cherenkov radiation emission angle as a function of $\phi$.}
\label{fig:leadtungstate}
\end{center}
\end{figure}

Finally, we perform calculations for lead tungstate crystals (PbWO$_4$) as for example used in the CMS calorimeter at the CERN LHC. At CMS, the lead tungstate light output is measured with Si avalanche photodiodes at a characteristic wavelength of 430 nm \cite{CMS_APD}. 
With a parameterization of the ordinary and extraordinary indices of refraction \cite{Chip98}, we get for this wavelength $n_o=2.3459$ and $n_e=2.2319$.
The results are are shown in Figure \ref{fig:leadtungstate}, and an asymmetry value of $R_t=0.9887$ is obtained, i.e.\ a small but possibly detectable effect.

These calculations could be tested in a simple experiment.  By choosing a crystal with a large value of abs$(R-1)$, like e.g. sodium nitrate, one would measure the total number of emitted photons as a function of $\chi$. 
One could also fix $\chi$ and measure the emission angle $\theta$ as a function of azimuthal angle $\phi$. 

It is even possible to choose a value of $\beta$ such that only parts of the Cherenkov cone would be filled. 
This measurement would be very sensitive to the $\beta$ value and constitute a very precise particle velocity test at a narrow velocity range. 

In all cases, the design of the crystal would have to take into account the fact that the Cherenkov emission angles are large.
This means that the rear end of the crystal must be constructed with a geometry such that total internal reflection is avoided. 

\section{Conclusions}

In anisotropic media, optical Cherenkov emission depends on the angle between the relativistic charged particle and the optical axis of the medium.  
In oriented ice crystals, the Cherenkov emission rate varies slightly, by 0.3\%, and the emission angle can vary by 0.4 degrees.
Since such crystals are found at the South Pole, this effect of anisotropic Cherenkov emission is important to understand for the neutrino experiments located on Antarctica.
However, the present results mean that such experiments, like IceCube, can safely neglect the effect of crystal anisotropy in their data analysis. 

Cherenkov radiation emission in materials where the difference between ordinary and extraordinary refractive indices is larger than for ice is impacted by the present results.
For the case of lead tungstate, the variation in angle is 1.4 degrees, and the intensity variation is slightly above one percent, both of which may be relevant for precise calorimetry. 
An experiment to test the accuracy of the formalism seems possible with relatively little effort.

SK acknowledges a useful conversation with Ryan Bay (UC Berkeley).  This work was supported in part by U.S. National Science Foundation under grants PHY-1307472  and the U.S. Department of Energy under contract number DE-AC-76SF00098.

\appendix*
\section{Appendix A}
This appendix presents the derivation of Eq. \ref{eq:N_e}. 
It uses an approach identical to the one presented in \cite{Muzicar,Delbart} and we use a notation similar to the one used in the latter, as in the rest of this paper.
We will just focus on the point where the two papers disagree and take the preceding calculations as given.
In agreement with both authors, we write
\begin{align}
\delta \left(\mathbf{\hat{u}\cdot \hat{r}} - \frac{v_p^{(e)}}{c\beta} \right) = \frac{\mathbf{\hat{u}\cdot \hat{r}}\delta \left( \cos(\theta) -\cos(\theta_{(e)})\right) }{\sqrt{(R-Q\cos^2(\phi))(P-Q\cos^2(\phi))}} \\
\mathbf{\hat{r}\cdot e^{(e)}} = \frac{\mathbf{\hat{u}\cdot \hat{r}}}{n_o^2\sqrt{1-u_1^2}} \left(r_1\beta^2n_o^2 \mathbf{\hat{u}\cdot \hat{r}}-u_1 \right),
\end{align}
with everything defined as in the main text.
These two expressions are the main ingredients in calculating the emission of extraordinary Cherenkov photons using Eq. \ref{eq:Ginzburg}.
We therefore evaluate
\begin{align}
\begin{split}
\mathbf{\hat{u}\cdot \hat{r}}\left( \mathbf{\hat{r}\cdot e}^{(e)}\right) ^2 &= \frac{\left(\mathbf{\hat{u}\cdot \hat{r}}\right)^4}{n_o^4 \left(1-u_1^2\right)} \left( r_1^2\beta^4n_o^4\left(\mathbf{\hat{u}\cdot \hat{r}}\right) + \frac{u_1^2}{\mathbf{\hat{u}\cdot \hat{r}}} - 2u_1r_1\beta^2n_o^2  \right) \\
&= \frac{\left(v_p^{(e)}\right)^4}{c^4n_o^2\beta^2 \left(1-u_1^2\right)} \left( r_1^2\beta^2n_o^2\left(\mathbf{\hat{u}\cdot \hat{r}}\right) + \frac{u_1^2}{\beta^2n_o^2\mathbf{\hat{u}\cdot \hat{r}}} - 2u_1r_1 \right) \\
&= \frac{\left(v_p^{(e)}\right)^4}{c^4n_o^2\beta^2 \left(1-u_1^2\right)} \left( r_1^2\beta^2n_o^2\left(\mathbf{\hat{u}\cdot \hat{r}}\right) + \frac{-(1 -u_1^2)+1}{\beta^2n_o^2\mathbf{\hat{u}\cdot \hat{r}}} - 2u_1r_1 \right)\\
&=  \frac{\left(v_p^{(e)}\right)^4}{c^4n_o^2\beta^2 }
\left( \frac{-1}{\beta^2n_o^2\mathbf{\hat{u}\cdot \hat{r}}}  +
\frac{1}{1-u_1^2}\left(r_1^2\beta^2n_o^2 \mathbf{\hat{u}\cdot \hat{r}}  + \frac{1}{\beta^2n_o^2\mathbf{\hat{u}\cdot \hat{r}}} -2u_1r_1 \right)\right) \\
&=\frac{\left(v_p^{(e)}\right)^4}{c^4n_o^2\beta^2 } \left( \frac{-1}{\beta^2n_o^2\mathbf{\hat{u}\cdot \hat{r}}}  +
\frac{1}{1-u_1^2}\left(\mathbf{\hat{u}\cdot \hat{r}} \left( r_1^2\beta^2n_o^2 -2\right)  + \frac{1}{\beta^2n_o^2\mathbf{\hat{u}\cdot \hat{r}}} -2u_2r_2 \right)\right),
\end{split}
\end{align}
where we have used the requirement $\mathbf{\hat{u}\cdot \hat{r}}= v_p^{(e)}/c\beta$ from the Dirac-delta function. 
The large parenthesis in this result is the same as the numerator in Eq. \ref{eq:N_e}.
Without giving any details on the calculation, both papers get this result wrong. 
It seems that the figures presented by Delbart et \textit{al.} are correct and that the paper just comes with some misprints. 
For instance, $\frac{r_1^2}{\beta^2n_o^2}$ is written rather than $r_1^2\beta^2n_o^2$. 
The confusion may have been caused by the notation in Muzicar where he defines $\beta_0=n_o^{-1}$.
Also the term $2u_2r_1$ is wrong since the two vector indices should be the same; either $1$ or $2$ depending on how the remainder of the equation is written. 
In the appendix of \cite{Delbart} the relation $\mathbf{\hat{u}\cdot \hat{r}} = (n_o\beta)^{-1}$ is used to integrate the result (\ref{eq:N_e}) despite that this relation only holds for the ordinary wave.
If the same relation is used on the above result, one gets an expression with a similar structure as the one presented in \cite{Delbart}. 
Whether one applies this relation changes the result by less than a percent. 
Finally, in the derivation there is also a minus sign missing between the two brackets in the expression for $A$ in their equation (4.5) - corresponding to (\ref{eq:A}) here. \\

In \cite{Muzicar}, there are two misprints: $r_1$ comes without an exponential factor of two and $n_e^2$ rather than $n_o^2$ is written in the first term of the numerator in Eq. \ref{eq:N_e}. 
The result is very similar to that of Delbart et. \textit{al}, so one could guess that the same relation which is true only for the ordinary wave has been applied.

\end{document}